\documentstyle[12pt]{article}
\begin{document}
\begin{center}
{\bf{Causality of the Brane Universe - OPERA and ICARUS}}

\vspace{0.5cm}

R.Parthasarathy{\footnote{e-mail address: sarathy@cmi.ac.in}} \\
Chennai Mathematical Institute \\
H1, SIPCOT IT Park \\
Siruseri 6003103 \\
India \\
\end{center}

\vspace{2.0cm}

{\noindent{\bf{Abstract}}}

\vspace{0.5cm}

The apparent violation of causality in the brane Universe can be avoided by taking the bulk spacetime
modelled by a 5-dimensional Kaluza theory with factorizable ansatz for the 5-dimensional metric whose components
do not depend on the fifth coordinate and with $G_{55}$ not a constant. The geodesic in the bulk does not correspond
to a free particle. The Kaluza scalar makes it non-inertial. The implications on the neutrino experiment is
that there is no superluminal propagation even after invoking sterile neutrinos.

\newpage

\vspace{0.5cm}

The idea that our Universe is a brane in some higher dimensional spacetime with the matter (physical) fields are 
confined in the brane and the gravitational fields live in the brane and higher dimensional spacetime, is gaining 
importance since the work of Randall and Sundrum [1]. In Ref.1, the brane is embedded in a 5-dimensional spacetime 
(with warped metric) without compactifying the fifth dimension. A possibility that matter in the brane may be 
connected almost instantaneously through the fifth dimension has been suggested by K\"{a}lbermann and Halevi [2].
The causal structure of the brane Universe with possible apparent violation of causality has been demonstrated by 
Ishihara [3]. This exotic scenario gained prominance recently to understand the results of the OPERA experiment 
on muon-neutrino beam [4] and many attempts were made to reconcile with the OPERA data [5,6,7,8] using the brane world 
approach. In these studies, the existence of 'stetrile neutrinos' is assumed. The 'active neutrinos' (in the 
Standard Model) propagate in the (3+1)-dimensional brane while the 'sterile neutrinos' (gauge singlets) are free to 
propagate in the extra-dimensional bulk as well as in the brane. Its geodesic between two points on the brane will 
include travel in bulk spacetime. By introducing a mixing of the sterile and active neutrinos, the studies show that 
the active neutrinos appear to be superluminal. There are other studies [9,10,11,12,13,14,15,16,17,18, partial list] 
on OPERA result not invoking brane world scenario and even suspecting the experiment [19].

\vspace{0.5cm}

Whether the results of the OPERA experiment are correct or not, the idea that there is a possible apparent 
violation of causality for the brane Universe merits further study. This becomes relevant in view of the recent 
experiment (ICARUS Collobaration) [20] that the time of flight difference between the speed of light and the 
arriving neutrino events is compatible with the simultaneous arrival of all events with equal speed. 

\vspace{0.5cm}

Briefly, Ishihara [3] considered the brane as a 4-dimensional intrinsically flat spacetime. Since the 
brane is embedded in the higher dimensional bulk spacetime, taken as 5-dimensional, there is the induced metric on 
the brane and an extrinsic curvature. The brane is curved extrinscally by its self gravity. It is then possible that 
a path appears in the bulk corresponding to two points on the brane. There will be two paths: a path confined on the 
brane with induced metric and extrinsic curvature and a path in the bulk which is free. If the information through the 
path in the bulk arrives earlier than the one through the path in the brane, then there is an apparent violation of 
causality. Ishihara [3] indeed showed that this happens when the brane is concave towards the bulk in the null 
direction. In other words, the fluctuations in the brane (via the extrinsic curvature) effectively increase the path 
length in the brane relative to the path length in the bulk. The 'on-brane geodesic' is longer than the 'bulk 
geodesic'. This conclusion is based on the realization that the geodesic in the bulk corresponds to a free particle. 

\vspace{0.5cm}

In this note, it will be shown that the geodesic in the bulk does not correspond to a free particle. As an 
illustrative  model for the bulk, we consider it a 5-dimensional spacetime as in the case of Kaluza-Klein theory. We 
assume a factorizable ansatz for the metric in the bulk for simplicity. The fifth coordinate is not taken to be 
compact. Further, the electromagnetism is switched off as photons live in the brane only. The 5-dimensional metric 
in the bulk is taken to be 
\begin{eqnarray}
G_{AB}\ =\ \left(\begin{array}{cc}
g_{\mu\nu}&0 \\
0&G_{55} \\
\end{array}\right)
&;&G^{AB}\ =\ \left(\begin{array}{cc}
g^{\mu\nu}&0\\
0&\frac{1}{G_{55}} \\
\end{array}\right),
\end{eqnarray}
where $A=\{\mu, 5\}$. We do not consider $G_{55}$ a constant. All the entries in the metric are taken to be 
independent of the fifth coordinate but depend on $x^{\mu}$. The above metric is taken as a model for the bulk. The 
brane world is the physical 4-dimensional spacetime with metric $g_{\mu\nu}(x)$. It is to be recalled that in the 
Kaluza theory which unifies gravity with electromagnetism, the $G_{55}$ component should not be a constant, though 
many authors considered this. With $G_{55}$ a constant, the Einstein equations in the 5-dimensional world become 
inconsistent. This important result was shown by Jordon [21] and Thiry [22] and reviewed by Overduin and Wesson [23]. 
In our case, with $G_{55}$ not a constant, the Einstein equations in the bulk for ${\tilde{R}}_{55}, {\tilde{R}}_{
\mu 5}, {\tilde{R}}_{\mu\nu}$ are consistent, in the absence of electromagnetism.

\vspace{0.5cm}

The geodesic equation in the 5-dimensional bulk spacetime is 
\begin{eqnarray}
\frac{d^2Z^A}{ds^2}+{\bigtriangleup}^A_{BC}\ \frac{dZ^B}{ds}\ \frac{dZ^C}{ds}&=&0,
\end{eqnarray}
where $Z^A\ =\ \{x^{\mu}, x^5\}$ and ${\bigtriangleup}^A_{BC}$'s are the 5-dimensional Christoffel connections;
${\bigtriangleup}^A_{BC}=\frac{1}{2}G^{AD}({\partial}_BG_{CD}+{\partial}_CG_{BD}-{\partial}_DG_{BC})$ with 
${\partial}_A=\frac{\partial}{\partial Z^A}$. We rewrite (2) as 
\begin{eqnarray}
\frac{d}{ds}\Big(G_{AB}\ \frac{dZ^B}{ds}\Big)-\frac{1}{2}({\partial}_AG_{CD})\ \frac{dZ^C}{ds}\ \frac{dZ^D}
{ds}&=&0.
\end{eqnarray}
The geodesics (2) or (3) are the geodesics for a free particle in the 5-dimensional bulk. Since the metric 
$G_{AB}$ in (1) are taken to be independent of the fifth coordinate, the $A=5$ part of (3) gives 
$G_{5B}\ \frac{dZ^B}{ds}$ a constant along the geodesic. We shall denote this constant by $a$. The $A=\mu$ part of 
(3) then becomes 
\begin{eqnarray}
\frac{d}{ds}\Big(g_{\mu\nu}\ \frac{dx^{\nu}}{ds}\Big)-\frac{1}{2}({\partial}_{\mu}g_{\nu\rho})\ \frac{dx^{\nu}}
{ds}\ \frac{dx^{\rho}}{ds}&=&\frac{1}{2}\frac{a^2}{ {G_{55}}^2}\ ({\partial}_{\mu}G_{55}),
\end{eqnarray}
which can be expressed as 
\begin{eqnarray}
\frac{d^2x^{\mu}}{ds^2}+{\Gamma}^{\mu}_{\rho\nu}\ \frac{dx^{\rho}}{ds}\ \frac{dx^{\nu}}{ds}&=&\frac{1}{2}
\frac{a^2}{ {G_{55}}^2}\ g^{\mu\lambda}({\partial}_{\lambda}G_{55}),
\end{eqnarray}
where ${\Gamma}^{\mu}_{\rho\nu}$ is the Christoffel connection for the metric $g_{\mu\nu}(x)$. This is the geodesic 
equation for a particle in the bulk 5-dimensional spacetime with indices $\mu,\nu,\rho,\lambda$ taking values 
$(0,1,2,3)$ as in the brane world. This describes the path appearing in the bulk corresponding to two points on the 
brane. Thus the path in the bulk, corresponding to two points on the brane, is not a 'free path'. The Kaluza 
scalar (the $G_{55}$ part of $G_{AB}$ in (1)) provides an additional force on the particle in the bulk. In this 
model, the 'on-brane geodesic' will not be longer than the 'bulk geodesic' as the geodesic in the bulk does not 
correspond to a free particle while the 'on-brane geodesic' corresponds to a free particle. 

\vspace{0.5cm}

If now, the fluctuations in the brane are taken into account by means of extrinsic curvature, then the geodesic on the 
brane has a contribution from the extrinsic curvature. When the brane is concave towards the bulk in the null 
direction, it is conceivable that the roles of the extrinsic curvature effects on the geodesic in the brane and that 
of the Kaluza scalar $G_{55}(x)$ on the geodesic in the bulk nicely balance so that the particles following the two 
paths, one on the brane and the other on the bulk, arrive eventually at the same time on the brane. This avoids the 
violation of causality in the brane Universe. 

\vspace{0.5cm}

Now analysing the reasoning of the explanation of the OPERA result by invoking sterile and active neutrinos with 
mixing, we see that the sterile neutrinos travelling in the bulk and the active neutrinos travelling in the 
brane could arrive at the same time on the brane, thereby negating the superluminal propagation of neutrinos.
It is gratifying that this is in agreement with the results of the recent ICARUS Collaboration [20] that the time 
of flight difference between the speed of light and the arriving neutrino events is compatible with the 
simulataneous arrival of all the events. 

\vspace{0.5cm}

The Kaluza's five dimensional spacetime description of the bulk (in which the Standard Model matter fields do not 
exist) - a model - with $G_{55}$ not a constant, is further examined in the light of constraints by Gubser [24]. 
Gubser [24] considered an extra-dimensional space with a non-factorizable ansatz for the metric and showed a 
possible violation of the null energy condition in extra dimensions. In our case, we have a factorizable metric in 
(1) and none of the components of $G_{AB}$ depend on the extra coordinate. Further, the Einstein equations in the 
bulk are ${\tilde{R}}_{AB}-\frac{1}{2}G_{AB}\tilde{R}\ =\ 0$ and so the null energy condition ${\tilde{R}}_{AB}
{\xi}^A{\xi}^B=0$ for ${\xi}^A$ any null vector is satisfied. So the constraint does not arise.  

\vspace{0.5cm}

To summarize: The apparent violation of causality in the brane Universe can be avoided by taking the bulk spacetime 
modelled by a 5-dimensional Kaluza theory with factorizable ansatz for the 5-dimensional metric whose components 
do not depend on the fifth coordinate and with $G_{55}$ not a constant. The geodesic in the bulk does not correspond 
to a free particle. The Kaluza scalar makes it non-inertial. The geodesic in the brane having extrinsic curvature
contribution makes it possible that the lengths of the geodesics in the bulk and on the brane can be made same, 
thereby avoiding the violation of causality in the brane Universe. The implications on the neutrino experiment is 
that there is no superluminal propagation even invoking sterile neutrinos. We have considered the fifth coordinate as 
non-compact. The concusions remain unaltered if the fifth coordinate is compactified to a circle. In this case, we 
need to retain the $n=0$ mode only as this corresponds to the lowest (vacuum) state.   

\vspace{1.5cm}

{\noindent{\bf{Acknowledgements}}}

\vspace{0.5cm}

Useful discussions with R.Jagannathan are acknowledged with thanks.

\vspace{0.5cm}

{\noindent{\bf{References}}}

\vspace{0.5cm}

\begin{enumerate}
\item L. Randall and R. Sundrum, Phys.Rev.Lett. {\bf{83}}, 4690 (1999). 
\item G. K\"{a}lbermann and H. Halevi, Nearness through an extra dimension, gr-qc/9810083.
\item H. Ishihara, Phys.Rev.Lett. {\bf{86}}, 381 (2001).
\item T. Adam et.al., (OPERA Collobaration) Measurement of the neutrino velocity with the OPERA detector in the 
      CNGS beam, hep-ex/1109.4897. 
\item G.F. Giudice, S. Sibiryakov and A. Strumia, Interpreting OPERA results on superluminal neutrino, 
      hep-ph/1109.5682.
\item S. Hannestad and M.S. Sloth, Apparent faster than light propagation from light sterile neutrinos,  
      hep-ph/1109.6282.
\item A. Nicolaidis, Neutrino Shortcuts in Spacetime, hep-ph/1109.6354.
\item D. Marfatia, H. P\"{a}s, S. Pakvasa and T.J. Weiler, Phys.Lett. {\bf{B707}}, 553 (2012), hep-ph/1112.0527.
\item A. Kehagias, Relativistic Superluminal Neutrinos, arXiv: 1109.6312.
\item N.D. Hari Dass, OPERA, SN1987a and energy dependence of superluminal neutrino velocity,  hep-ph/1110.0351.
\item I.Ya. Arefeva and I.V. Volovich, Superluminal Dark Neutrinos, hep-ph/1110.0456.
\item M. De Sanctis, Wave packet distortion and superluminal neutrinos, hep-ph/1110.3071.
\item H. Bergeron, About Statistical Questions involved in the Data Analysis of the OPERA Experiment, 
      arXiv: 1110.5275.
\item D. Indumathi, R.K. Kaul, M.V.N. Murthy and G. Rajasekaran, Group Velocity of Neutrino Waves,  arXiv: 1110.5453.
\item N. Nakanishi, An interpretation of 'Superluminal Neutrino' compatible with Relativity in the framework of 
      Standard Model, hep-ph/1111.1760.
\item J.D. Franson, Reduced photon velocities in the OPERA Neutrino Experiment and Supernova 1987a, hep-ph/1111.6986.
\item E. Ciuffoli, J. Evslin, X. Bi and X. Zhang, Density-Dependent Neutrino Dispersion Relation for OPERA?, 
      hep-ph/1112.3551.
\item G. Amelino-Camelia, Third road to the OPERA: a tunnel after all?, hep-ph/1201.6496.   
\item A.G. Cohen and S.L. Glashow, New Constraints on neutrino velocities, hep-ph/1109.6562. 
\item M. Antonello et.al., (ICARUS Collaboration), Measurement of the neutrino velocity with the ICARUS detector at 
      the CNGS beam,  arXiv: 1203.3433. 
\item P. Jordan, Ann.Phys. (Leipzig) {\bf{1}}, 219 (1947).
\item Y. Thiry, Comptes Rendus Acad.Sci. (Paris) {\bf{226}}, 216 (1948).
\item J.M. Overduin and P. Wesson, Phys.Rep. {\bf{283}}, 303 (1997); gr-qc/9805018. 
\item S.S. Gubser, Superluminal neutrinos and extra dimensions: Constraints from the null energy condition. hep-th/1109.5687. 
\end{enumerate}

\end{document}